\documentclass[preprint,showpacs,preprintnumbers,amsmath,amssymb]{revtex4}

\usepackage{graphicx}
\usepackage{dcolumn}
\usepackage{bm}
\usepackage{epsfig}

\def\e{\begin{equation}}
\def\f{\end{equation}}
\def\=#1{\overline{\overline #1}}

\def\-#1{{\bf #1}}
\def\.{\cdot}
\def\l#1{\label{eq:#1}}
\def\r#1{(\ref{eq:#1})}

\def\dyad#1{\overline{\overline {\bf{#1}} }}

\begin{document}

\title{Broadband Negative Refraction with a Crossed Wire Mesh}

\author{M\'ario G. Silveirinha}
 \altaffiliation[]{Electronic address: mario.silveirinha@co.it.pt}
 \affiliation{University of Coimbra, Department of Electrical
  Engineering-Instituto de Telecomunica\c{c}\~{o}es, 3030 Coimbra, Portugal}

\date{\today}

\begin{abstract}
It is demonstrated that a structured material formed by nonconnected
crossed metallic wires may enable negative refraction over a wide
frequency range. This phenomenon is a consequence of the anomalous
dispersion characteristics of the material, particularly of the fact
that the isofrequency contours are hyperbolic. These properties rely
on the nonlocal response of the crossed wire mesh, and establish a
different paradigm for obtaining negative refraction without
left-handed materials.
\end{abstract}

 \maketitle

\section{Introduction}

Negative refraction has undoubtedly been in the spotlight in recent
years. Much of the interest in this fascinating phenomenon is caused
by the fact it contradicts our experience that light is bent by
common materials (e.g. glass) in a such a way that the projections
of the incident and transmitted rays along the interface are
oriented in the same direction. It was first predicted by Veselago
\cite{Veselago} forty years ago that materials with simultaneously
negative permittivity and permeability (``left-handed'' materials)
may enable negative refraction. However, such exotic materials were
unknown at the time, and only some years ago they were made
available in the form of metamaterials \cite{SmithDNG}. Shortly
after this finding negative refraction was demonstrated at
microwaves \cite{Shelby}, but, since structuring materials in the
nanoscale is still a true challenge, only very recently negative
refraction was finally revealed at optical frequencies using a truly
three-dimensional metamaterial \cite{ZhangNature}.

It is well-known that negative refraction may be obtained without
left-handed materials. For example, one possibility is to use
indefinite anisotropic materials, for which one component of the
permittivity is negative \cite{Smithindef, FanNegRef, Hoffman,
ZhangScience, ZhangOE}. Another possibility is to engineer the
dispersion characteristic of photonic crystals \cite{NotomiNegRef,
Cubukcu}. Here, we demonstrate a distinctively different route to
obtain all-angle broadband negative refraction using a spatially
dispersive material formed by a crossed wire mesh. It is shown that
due to the strongly nonlocal response of the structured material the
dispersion contours of the propagating mode consist of two
hyperbolas, and that this property makes the group velocity (energy
flow) to be refracted in an anomalous manner at an interface with
air.
\begin{figure}[th] \centering
\epsfig{file=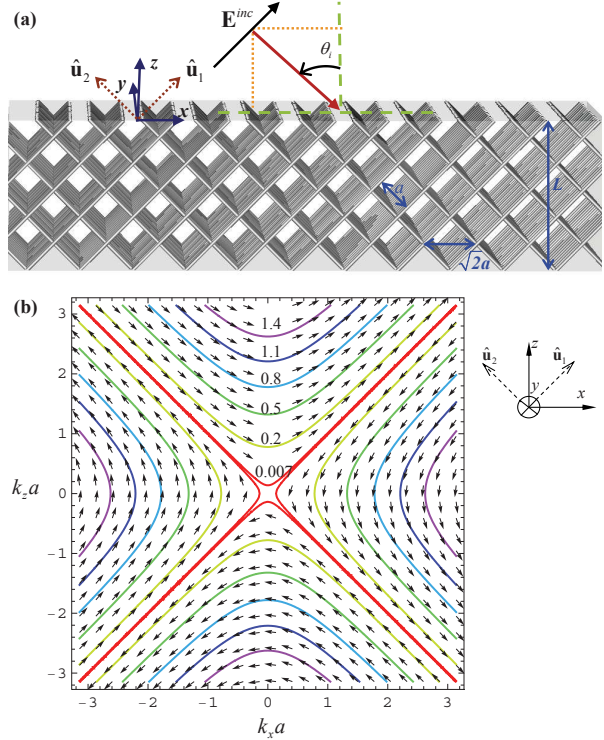, width=8.0cm} \caption{(Color online)
Panel (a): Geometry of the metamaterial: two nonconnected wire
meshes are oriented at right-angles. The distance (along $y$)
between perpendicular adjacent wires is $a/2$. Panel (b):
Isofrequency contours of the fundamental plane wave mode for
propagation in the $xoz$ plane with electric field in the same
plane. The text insets indicate the value of the normalized
frequency $\omega a/c$. The arrows show the orientation of the
electric field associated with a given isofrequency contour.}
\label{Isofreq}
\end{figure}

The material considered here is a crossed wire mesh of nonconnected
metallic wires with radius $r_w$. The orientation of the two arrays
of wires is determined by the perpendicular unit vectors ${\bf{\hat
u}}_1$ and ${\bf{\hat u}}_2$ [Fig. \ref{Isofreq}a]. It was
demonstrated in our previous work \cite{MarioEVL} that such material
may be characterized by an anomalously high $positive$ index of
refraction in the long wavelength limit, and that such property may
enable the propagation of very subwavelength guided modes, as
demonstrated experimentally in \cite{EVLexp}. Recently, we have
shown that the resonant excitation of such guided modes may enable
subwavelength imaging \cite{XwiresLens}.

Evidently, the crossed wire mesh has an anisotropic electromagnetic
response. The results of our previous studies \cite{MarioEVL,
EVLexp, XwiresLens} assumed electromagnetic propagation in a plane
normal ($yoz$) to the planes of wires. Here, we demonstrate that for
propagation along a direction parallel to the planes of wires
($xoz$) the propagation properties may be dramatically different,
and may enable negative refraction. To demonstrate these potentials
first we will characterize the isofrequency contours of the bulk
metamaterial. As reported in \cite{MarioEVL, Silv_MTT_3DWires,
Constantin_WM, IgorWM2D}, in the long wavelength limit the crossed
wire mesh may be characterized by the dielectric function
$\dyad{\varepsilon} = {\varepsilon _{h} \bf{\hat u}}_y {\bf{\hat
u}}_y + \varepsilon _{11} {\bf{\hat u}}_1 {\bf{\hat u}}_1 +
\varepsilon _{22} {\bf{\hat u}}_2 {\bf{\hat u}}_2 $ ($\varepsilon
_{h}$ is the relative permittivity of the host medium), with
\e\varepsilon _{ii} \left( {\omega ,k_i } \right) = \varepsilon _h
\left( {1 - \frac{{\beta _p^2 }}{{\left( {\omega /c} \right)^2
\varepsilon _h  - k_i^2 }}} \right), \quad i=1,2 \l{epsWM} \f where
$ \beta _p = \left[ {2\pi /\left( {\ln \left( {a/2\pi r_w } \right)
+ 0.5275} \right)} \right]^{1/2} /a$, $c$ is the speed of light in
vacuum, ${\bf{k}} = \left( {k_x ,k_y ,k_z } \right)$ is the wave
vector, $k_i = {\bf{k}}{\bf{.\hat u}}_i $, and $a$ is the distance
between adjacent parallel wires. For simplicity, it is assumed that
the wires are perfect conductors. The effect of metallic loss can be
easily accounted for using the more general formulae of Ref.
\cite{MarioEVL}, and is relatively small provided the radius of the
wires $r_w$ is larger than the skin depth of the metal
\cite{MarioEVL}, which may be verified through the infrared domain.
It is convenient to choose the system of Cartesian axes so that
${\bf{\hat u}}_1 = {{\left( {1,0,1} \right)}/{\sqrt 2 }}$ and
${\bf{\hat u}}_2 = {{\left( {-1,0,1} \right)}/{\sqrt 2 }}$ [see Fig.
\ref{Isofreq}a].

Consider the case in which the electromagnetic wave propagates in
the $xoz$ plane, $k_y=0$, with magnetic field polarized along $y$
(the time variation $e^{-i \omega t}$ is suppressed): \e {\bf{H}} =
H_0 e^{ i {\bf{k}}{\bf{.r}}} {\bf{\hat u}}_y ,\quad {\bf{k}} =
\left( {k_x ,0,k_z } \right)\l{Hfield}\f The corresponding electric
field is given by: \e {\bf{E}} = \frac{{H_0 }}{{\omega \varepsilon
_0 }}\left( {\frac{k_2}{{\varepsilon _{11} }}{\bf{\hat u}}_1  -
\frac{k_1}{{\varepsilon _{22} }}{\bf{\hat u}}_2 } \right) e^{
i{\bf{k}}{\bf{.r}}}. \f Using the Maxwell's Equations it is
straightforward to verify that the wave vector must follow the
dispersion characteristic: \e \frac{{k_1^2 }}{{k^2  - \left( {\omega
/c} \right)^2 \varepsilon _{11} }} + \frac{{k_2^2 }}{{k^2  - \left(
{\omega /c} \right)^2 \varepsilon _{22} }} = 1 \l{dispEq}\f which,
using Eq. \r{epsWM}, may be reduced to a polynomial equation of
third degree in $\beta^2$ with $\beta=\sqrt{\varepsilon_h}\omega/c$:
\begin{eqnarray}
&&- 4\beta ^6  + 8\beta ^4 \left( {\beta _p^2  + k_x^2  + k_z^2 }
\right) - \nonumber \\
&&\beta ^2 \left( {  4\beta _p^4  + 8\beta _p^2 k_x^2  +
8\beta _p^2 k_z^2  + 6k_x^2 k_z^2  + 5k_x^4  + 5k_z^4 } \right) + \nonumber \\
&&\left( {k_x^2  - k_z^2 } \right)^2 \left( {2\beta _p^2  + k_x^2  +
k_z^2 } \right) = 0 \l{dispEQ}
\end{eqnarray}
It may be verified that when the operating wavenumber is much
smaller than the plasma frequency, $\beta \ll \beta_p$, there is a
unique positive solution for $\beta^2$. This means that for low
frequencies and for a given ${\bf{k}} = \left( {k_x ,0,k_z }
\right)$ there is always a propagating mode with electric field in
the $xoz$ plane. The isofrequency contours of this propagating mode
are represented in Fig. \ref{Isofreq}b for a material with
$r_w=0.05a$ formed by metallic wires standing in air
($\varepsilon_h=1$). As seen in Fig. \ref{Isofreq}b and also
reported in Ref. \cite{Silv_MTT_3DWires, Constantin_WM}, for a fixed
frequency the isofrequency contours consist of two hyperbolas with
asymptotes running along the directions ${\bf{\hat u}}_1$ and
${\bf{\hat u}}_2$. These hyperbolic contours resemble in part the
isofrequency contours of an indefinite material, however there is an
important difference: the isofrequency contour of an indefinite
material consists of a single hyperbola \cite{Smithindef}. In Fig.
\ref{Isofreq}b we have also represented the electric field vector
lines, which are qualitatively similar to those in an indefinite
material.
\begin{figure}[th] \centering
\epsfig{file=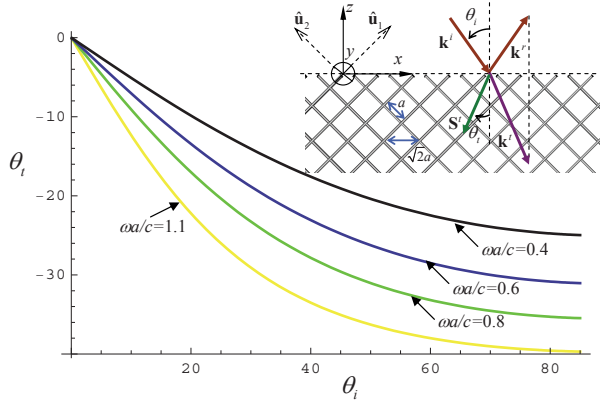, width=8.0cm} \caption{(Color online)
Angle of transmission of the energy flow (Poynting vector) as a
function of the angle of incidence, for different frequencies of
operation. The radius is $r_w=0.05a$ and the host permittivity is
$\varepsilon_h=1$. The inset represents the geometry of the problem
showing the incident, reflected and refracted waves. Notice that the
wave vector and the Poynting vector are not parallel in the
metamaterial.} \label{refraction}
\end{figure}

It is well known that hyperbolic contours may enable negative
refraction at an interface with air \cite{Smithindef, FanNegRef,
ZhangScience, ZhangOE}. In order to investigate this possibility, we
consider the geometry shown in the inset of Fig. \ref{refraction},
which shows an incoming plane wave illuminating the metamaterial
along the direction $\theta_i$. The angle of refraction $\theta_t$
for the energy flow (determined by the Poynting vector) can be
calculated using the fact that the projection of the wave vector
onto the interface is preserved \cite{NotomiNegRef}. Thus, for a
given frequency $\omega$ and angle of incidence $\theta_i$, the wave
vector associated with the transmitted wave is of the form
${\bf{k}}^t=\left( \omega/c\, \sin \theta_i ,0,k_z^t \right)$, where
$k_z^t$ is calculated using the dispersion equation \r{dispEQ}
\footnote{For frequencies larger than the plasma frequency, it may
be possible that a plane wave propagating in air excites two
propagating modes in the metamaterial, and thus originates two
refracted beams. However in the long wavelength regime considered
here, $\beta  \ll \beta _p$, only a single propagating mode can be
excited, and thus there is a single refracted beam.}. The Poynting
vector is normal to the associated isofrequency contour
\cite{Agranovitch}, \footnote{In nonlocal media the Poynting vector
can be explicitly calculated as, ${\bf{S}} = {\bf{S}}^{\left( 0
\right)}  + {\bf{S}}^{\left( 1 \right)}$, where
 ${\bf{S}}^{\left( 0 \right)}  = \frac{1}{2}{\mathop{\rm Re}\nolimits} \left\{ {{\bf{E}} \times {\bf{H}}^* }
 \right\}$ and the ``high frequency'' component $
{\bf{S}}^{\left( 1 \right)} $ is such that $ S_l^{\left( 1 \right)}
= - \frac{{\omega \varepsilon _0 }}{4}{\mathop{\rm Re}\nolimits}
\left\{ {{\bf{E}}^* {\bf{.}}\frac{{\partial \overline{\overline
\varepsilon } }}{{\partial k_l }}{\bf{.E}}} \right\}, \, (l=1,2,3) $
\cite{IgorWM2D, Agranovitch}. It is demonstrated in
\cite{Agranovitch} (p. 65-67) that the Poynting vector is given by
the product of the energy density and the group velocity.}. As in
indefinite media, the angle between the wave vector and the Poynting
vector in the metamaterial is acute.

In Fig. \ref{refraction} we show the angle $\theta_t$ as a function
of the angle of incidence $\theta_i$, for different frequencies of
operation. Consistent with hyperbolic shape of the isofrequency
contours, the angle of transmission $\theta_t$ is negative, i.e. the
wave group velocity suffers, indeed, negative refraction at the
interface of the crossed wire mesh with air. The results of Fig.
\ref{refraction} indicate that this phenomenon (unlike in photonic
crystals and left-handed materials) is very broadband, being
observed for a wide range of frequencies. Notice that at the
considered frequencies the electrical size of the unity cell of the
crossed wire mesh is small, $a \ll \lambda$, as required in order
that homogenization theory can be applied. The emergence of negative
refraction in the crossed wire mesh has a simple physical
justification. Indeed, consider the scenario depicted in
Fig.\ref{Isofreq}a, which shows the incoming wave illuminating the
metamaterial along a direction such that the incoming electric field
is nearly parallel to the direction ${\bf{\hat u}}_1$. In these
circumstances, it is clear from the topology of the metamaterial
that the incoming wave will interact mainly with the set of wires
directed along ${\bf{\hat u}}_1$. But since these wires are tilted
by $-45^o$ with respect to the interface, it is apparent that as the
wave propagates in the crossed wire mesh it suffers a negative
spatial shift, or in other words, since the dominant path of
propagation is expected to be along the wires parallel to ${\bf{\hat
u}}_1$, the group velocity suffers negative refraction. Thus, in a
certain sense, each set of wires behaves as a waveguide, and the
polarization of the incoming wave controls which set of wires is
``activated'' and which set of wires is ``dormant''. This heuristic
interpretation is of course only a very rough description of the
complex wave interaction between the crossed wires, but enables one
to visualize and relate the negative refraction to the
microstructure of the material.
\begin{figure}[t] \centering
\epsfig{file=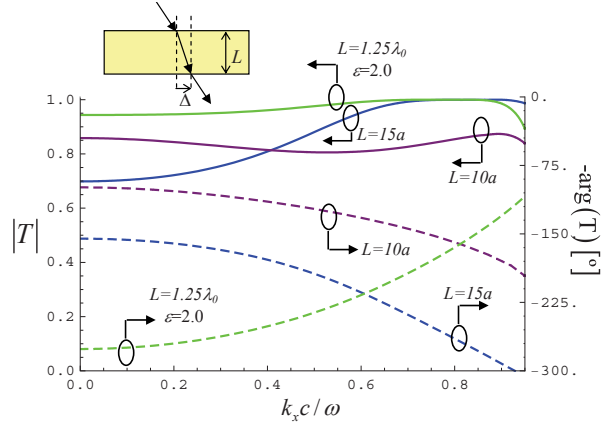, width=8.0cm} \caption{(Color online)
Transmission coefficient as a function of normalized $k_x$ for a
fixed frequency and different $L$. Solid lines: magnitude; Dashed
lines: phase. The lattice constant is such that $\omega a/c =0.6$,
and the radius of the wires is $r_w=0.05a$. The curves associated
with $L=1.25 \lambda_0$ were calculated for a regular dielectric
slab with permittivity $\varepsilon=2$.} \label{spatshift}
\end{figure}

In order to confirm the homogenization results, we used the full
wave commercial simulator CST Microwave $\rm{Studio^{TM}}$ to
calculate the transmission coefficient $T$ of a crossed wire mesh
slab under plane wave incidence, with magnetic field along the
$y$-direction [Fig. \ref{spatshift}]. Specifically, we have
calculated $T$ as a function of $k_x$ ($x$- component of the wave
vector of the incoming wave) for a normalized frequency of operation
$\omega a/c = 0.6$. Notice that $k_x$ is determined by the angle of
incidence: $k_x = \omega/c \, \sin \theta_i$. It can be seen in Fig.
\ref{spatshift} that the crossed wire mesh is relatively transparent
to radiation for all incident angles, and that $\left| T \right|$
may be close to unity. This property is observed over a very wide
frequency range (not reported here for brevity); in general, the
transmissivity tends to improve with increasing frequency. These
results indicate that the crossed wire mesh may be well-matched to
free-space.

It is interesting to analyze the variation of the phase of $T$ with
$k_x$. For convenience, let us write $T = \left| T \right|e^{ -
i{\kern 1pt} \phi }$, where $\phi  =  - \arg T $. It can be seen in
Fig. \ref{spatshift} that $\phi$ is a decreasing function of $k_x$.
Interestingly, such behavior is completely different from that in a
conventional dielectric slab with positive permittivity (green
dashed line in Fig. \ref{spatshift}), for which $\phi$ increases
with $k_x$. Indeed, as proven next, the slope of $\phi$ is equal to
the spatial shift $\Delta$ suffered by an incoming beam when it
crosses the slab [inset of Fig. \ref{spatshift}]. In fact, suppose
that an incoming beam (e.g. with Gaussian profile) impinges on the
planar slab. Suppose that the magnetic field at the input interface
is $H_y= H_y^i(x)$. Using Fourier theory the incoming beam can be
written as a superposition of plane waves, as $H_y^i \left( x
\right) = \int {\tilde H_y \left( {k_x } \right)e^{ik_x x} dk_x }$.
Evidently, $T=T(k_x)$ may be regarded as the transfer function of
the slab. This means that the magnetic field at the output is such
that: $H_y^o \left( x \right) = \int {\tilde H_y \left( {k_x }
\right)T\left( {k_x } \right)e^{ik_x x} dk_x }$. But, if the spatial
spectrum of the incoming beam is highly concentrated at the wave
number $k_x^0=\omega/c \sin \theta_i$, i.e. if the beam is a
quasi-plane wave propagating along the direction $\theta_i$, a
straightforward analysis (similar to the one used to define group
velocity) shows that the magnetic field at the output plane is such
that: $H_y^o \left( x \right) \approx T\left( {k_x^0 }
\right)e^{i\Delta \,k_x^0 } H_y^i \left( {x - \Delta } \right)$,
where $\Delta  = d\phi /dk_x$. Therefore, apart from a transmission
coefficient, the field at the output plane differs from the field at
the input plane from a spatial shift $\Delta$, which is completely
determined by the slope of the phase. This analysis is completely
general and is valid for an arbitrary material slab. It provides a
simple criterion to test the emergence of negative refraction in
metamaterials, by testing if $\Delta$ is positive or negative.
Applying the proposed theory to the crossed wire metamaterial, we
see from Fig. \ref{spatshift} that the slope of $\phi$ is negative,
and thus it follows that the spatial shift $\Delta$ is also
negative. This is a direct proof of the emergence of negative
refraction in the metamaterial, and fully supports the
homogenization results. In particular, for the curve associated with
$L=15a$ ($L=10a$) we have numerically calculated (for incidence
along $\theta_i=45^o$), $\Delta = d\phi /dk_x = -0.5L$ ($-0.39L$),
which indicates that the transmission angle is $\theta _t = \tan ^{
- 1} \Delta /L = -27^o$ ($\theta _t  =-21^o$). These values concur
well the value predicted by homogenization, $\theta _t  =-25^o$
[Fig. \ref{refraction}].

\begin{figure*}[t] \centering
\includegraphics[width=17cm]{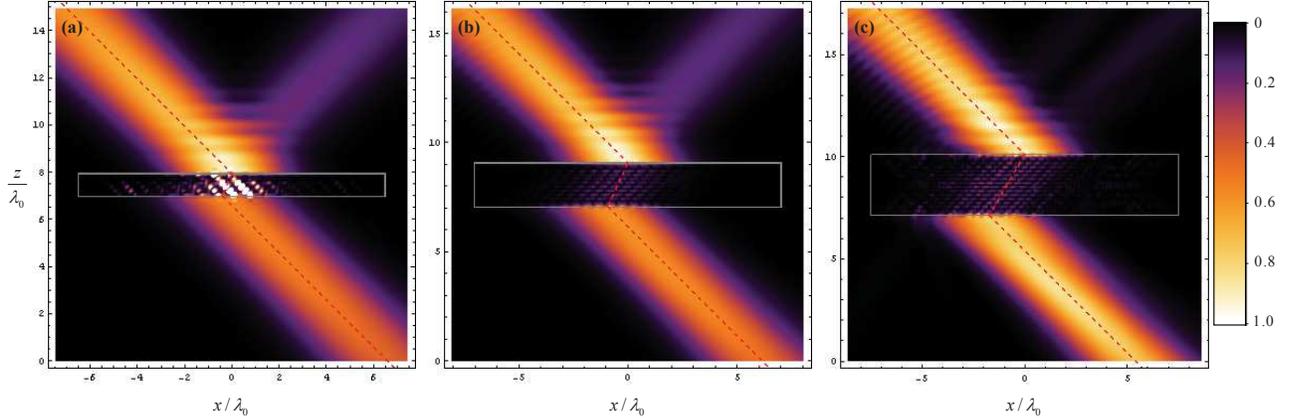}
\caption{(Color online) Normalized $\left| {\bf{E}} \right|^2$ in
the vicinity of the crossed wire mesh calculated by solving the
Maxwell's equations using the Method of Moments. The incoming wave
has a Gaussian profile. Both the metamaterial slab and the incoming
beam are periodic along $y$. The wires lie in planes parallel to the
$xoz$ plane and have radius $R=0.05a$. Each plane of wires contains
90 wires. The frequency of operation is $\omega a/c = 0.6$. Panel
(a): $L=10.0a$; Panel (b): $L=21.0a$. Panel (c): $L=31.0a$.}
\label{NegRefrac}
\end{figure*}

To demonstrate in a conclusive manner the appearance of negative
refraction, we used the Method of Moments (MoM) to simulate
numerically the response of a finite width metamaterial slab
illuminated by an incoming beam with Gaussian profile. It is assumed
in the simulation that both the structured material and the incoming
beam are periodic along the $y$-direction, with period equal to $a$.
The MoM numerically solves  the Maxwell's Equations and takes into
account all the details of the microstructure of the crossed wire
mesh, yielding the exact solution of the problem, apart from
``numerical noise''. It is assumed in the simulations that each
plane of inclusions (parallel to the $xoz$ plane) is formed by 90
wires with radius $R=0.05a$. As depicted in Fig. \ref{Isofreq}a, the
wires in alternate planes are perpendicular. The Gaussian beam
illuminates the slab along the direction $\theta_i=45^o$, and has a
beam waist equal to $2 w_0 = 4.0 \lambda$ at the normalized
frequency of operation $\omega a/c = 0.6$ (i.e. $a=0.1 \lambda$).
The width of the slab along $x$ is approximately $W=90 a \sqrt{2}
\approx 12 \lambda$. In Fig. \ref{NegRefrac} we represent the
calculated squared amplitude of the electric field (which is roughly
proportional to the beam intensity) in the vicinity of the crossed
wire mesh for different thicknesses of the slab. The dashed line
represents the propagation path (maximum of the electric field
amplitude), and clearly shows that the incoming wave is bent with a
negative transmission angle at the interfaces with air. The angle
$\theta_t$ calculated directly from the spatial shift $\Delta$ is
$\theta_t=-22^o$, $\theta_t=-25^o$, and $\theta_t=-30^o$ for panels
(a), (b), and (c) of Fig. \ref{NegRefrac}, respectively. These
values match relatively well the theoretical value $\theta _t
=-25^o$ [Fig. \ref{refraction}].

It is interesting to note that the electric field amplitude is
significantly lower in the metamaterial slab, as compared to the air
regions, particularly in panels (b) and (c) of Fig. \ref{NegRefrac}.
One of the reasons is that the field distribution of panels (b) and
(c) was calculated at a plane equidistant from two adjacent wire
planes, whereas the field distribution of panel (a) was evaluated on
a plane with wires. The second reason is that the wave impedance
$\eta$ in the crossed wire mesh is lower than in free-space
\cite{MarioEVL}, and thus, since beam intensity is roughly $ \left|
{\bf{E}} \right|^2 /2\eta $, the conservation of energy requires
that the squared amplitude of the electric field is lower in the
metamaterial. Despite the difference between the impedance in the
structured material and free-space, it is evident from the
simulations that the level of reflections is relatively weak.

In conclusion, we have demonstrated that a crossed wire mesh may
enable negative refraction over a wide frequency band. The described
phenomenon does not rely on a resonance of the inclusions, and due
to this reason the effect of loss is expected to be relatively
small, particularly when the radius of the wires is smaller than the
skin depth of metal \cite{MarioEVL}. A block of the considered
material enables negative refraction either if the interface is
normal to $x$- or to the $z$- direction. In this regard, the
response of the metamaterial is fundamentally different from that of
a conventional indefinite anisotropic material, which only yields
negative refraction when the interface is normal to the principal
axis along which the permittivity is negative. The described results
illustrate the richness of the physics of nonlocal materials. This
work is supported in part by Funda\c{c}\~ao para a Ci\^encia e a
Tecnologia under project PDTC/EEA-TEL/71819/2006.

\bibliography{refs}
\end{document}